\documentclass{ceab}   

\usepackage{epsfig}     
\usepackage{graphicx}   

\usepackage{ceabbib}     
\usepackage[T1]{fontenc}

\begin{document}

\title{Coronal mass ejections associated with LDE flares of slow rise phase}

\author{U. B\c ak-St\c e\' slicka, S. Ko\l oma\'nski and T. Mrozek
\vspace{2mm}\\
\it Astronomical Institute, University of Wroc{\l}aw \\ 
\it ul. Kopernika 11, 51-622 Wroc{\l}aw, Poland\\}

\maketitle

\begin{abstract}
It is well known, that there is temporal relationship between CMEs and associated flares. The duration of the acceleration phase is related to the duration of the rise phase of a flare. We investigated CMEs associated with long duration flares of slow rise phase (slow LDE). These CMEs are characterized by high velocity ($v>1000$ km/s) during the propagation phase and low average acceleration during the main, prolongated, acceleration phase. The CMEs are accelerated until the height $> 5 R_\odot$, which is higher value than in typical impulsive CMEs. CMEs associated with slow LDEs have characteristics of both classes of CMEs, i.e. CMEs associated with flares and CMEs associated with eruptive prominences.

\end{abstract}


\section{Introduction}
The first Coronal Mass Ejection (CME) was clearly identified by R. Tousey in 1971 \citep{tousey1973} in the {\em Orbiting Solar Observatory-7} observations. Later on CMEs were observed by {\em Skylab} \citep{gosling1974} and since that time have become one of the most interesting topics in solar physics. It is well known that CMEs are associated with flares and eruptive prominences \citep{munro1979}.
 
Basing on small samples of events \cite{gosling1976} and \cite{macqueen1983} showed that the flare-associated CMEs are characterized by high velocities (approximately constant) and rapid acceleration. On the other hand CMEs associated with the prominences are accelerated over larger height-range and attain lower velocities than flare-associated CMEs. Using {\em SoHO}/LASCO data \cite{sheeley1999} separated CMEs into two classes: 1) gradual CMEs with lower speed ($400$--$600$ km/s), associated with eruptive prominences; 2) impulsive CMEs with higher speed ($>750$ km/s), associated with flares and showing clear evidence of deceleration. The existence of two classes of CMEs was confirmed by many authors \citep{stcyr1999,andrews2001,moon2002}. However, there are also observations that not confirm differences between the two types of CMEs \citep{dere1999}. \cite{vrsnak2005} presented statistical analysis of 545 flare-associated CMEs and 104 non-flare CMEs. They demonstrated that both data sets show very similar characteristics: in both samples significant fraction of CMEs was accelerated or decelerated and both samples include a comparable ratio of fast and slow CMEs. This directly opposes the concept of two distinct (flare/non-flare) classes of CMEs. On average, mean CME velocities gradually decrease, following the sequence: strong-flare events $\rightarrow$ weak-flare events $\rightarrow$ non-flare events with eruptive prominences $\rightarrow$ no-flare/no-prominence events.

\cite{zhang2001} investigated the temporal relationship between CMEs and associated solar flares. They found that kinematic evolution of CMEs can be described in three-phase scenario: the initiation phase, the impulsive acceleration phase, and the propagation phase. The impulsive acceleration phase is often synchronized with the rising phase of the associated soft X-ray flare (\cite{zhang2001,zhang2004,maricic2007} and references therein), i.e., during the flare impulsive phase (\cite{temmer2008,temmer2010} and references therein). Typically, CMEs are accelerated over short distance range, usually below the occulting disc of white-light space-borne coronagraphs. However, sometimes they accelerate up to heights of of several solar radii, in some cases even until the height of 7 $R_{\odot}$ \citep{zhang2001,zhang2004,vrsnak2001}.

During the acceleration phase, a CME has an acceleration of a few hundred m/s$^{2}$ \citep{wood1999,zhang2001}. The value of the acceleration of a CME depends on a duration of the main acceleration phase \citep{zhang2006,vrsnak2007}. A fast CME ($v > 1000$ km/s) can be strongly accelerated over short time, or weakly accelerated over an extended time interval of several hours \citep{zhang2004,vrsnak2007}. Since CMEs are associated with flares, a duration of the acceleration phase is related to a duration of the rise phase of a flare. CMEs associated with flares of slow rising phase should be accelerated slowly but for a long time.

Here, we investigate the kinematic evolution of CMEs associated with flares of slow rising phase. These CMEs were accelerated as long as SXR flux of associated flares increased. During the main acceleration phase CMEs reached heights $> 5 R_\odot$, which is a higher value than for a typical CME. These CMEs are characterized by a high velocity during the propagation phase and a low average acceleration during the main, prolongated, acceleration phase. Height-time profiles for the CMEs are similar to profiles of CMEs associated with eruptive prominences, which means that those CMEs may have characteristics both groups of CMEs, i.e. CMEs associated with flares and CMEs associated with eruptive prominences.

\section{Observations}
We selected five CMEs associated with slow LDEs. We were interested only in near-the limb events to avoid projection effects. Basic information about associated slow LDEs are given in Table \ref{flares}. We determined heights of the CMEs (in the plane of the sky) using running-differences images from the {\em SoHO}/EIT, MLSO/MK4\footnote{http://mlso.hao.ucar.edu/cgi-bin/mlso\_data.cgi} and {\em SoHO}/LASCO C2 and C3 instruments. We fitted second-degree polynomial to the points on the height-time profile to obtain average acceleration during the main acceleration phase. Time range of data used for quadratic fitting estimates the acceleration phase duration. We used linear fit to obtain average velocity during the propagation phase. Here, we present two events in detail. The results for all the analysed CMEs are shown in Table \ref{CMEs}. 

\begin{table}[!t]
\begin{center}
	\caption{Data for flares associated with the CMEs}
	\vspace{3mm}
	\label{flares}
	\begin{footnotesize}
		\begin{tabular}{|c|c|c|c|c|c|c|}
\hline
&Flare&Flare& Rise phase&{\em GOES}&&\\
Date&onset&peak&duration&class&NOAA&Location\\
&[UT]&[UT]&[min]&&&\\
\hline
21-May-02&23:14&00:30&76&C9.7&9948&S25W65\\
\hline
28-Jul-04&02:32&06:09&217&C4.4&10652&N03W66\\
\hline
29-Jul-04&11:42&13:04&82&C2.1&10652&N03W84\\
\hline
31-Jul-04&05:16&06:57&101&C8.4&10652&N03W105\\
\hline
06-Sep-05&19:32&22:02&150&M1.4&10808&S12E96\\
\hline
	\end{tabular}
	\end{footnotesize}
\end{center}
\end{table}

\section{Detailed analysis of two events}
\begin{figure}[!t]
\begin{center}
\includegraphics[width=6.8cm]{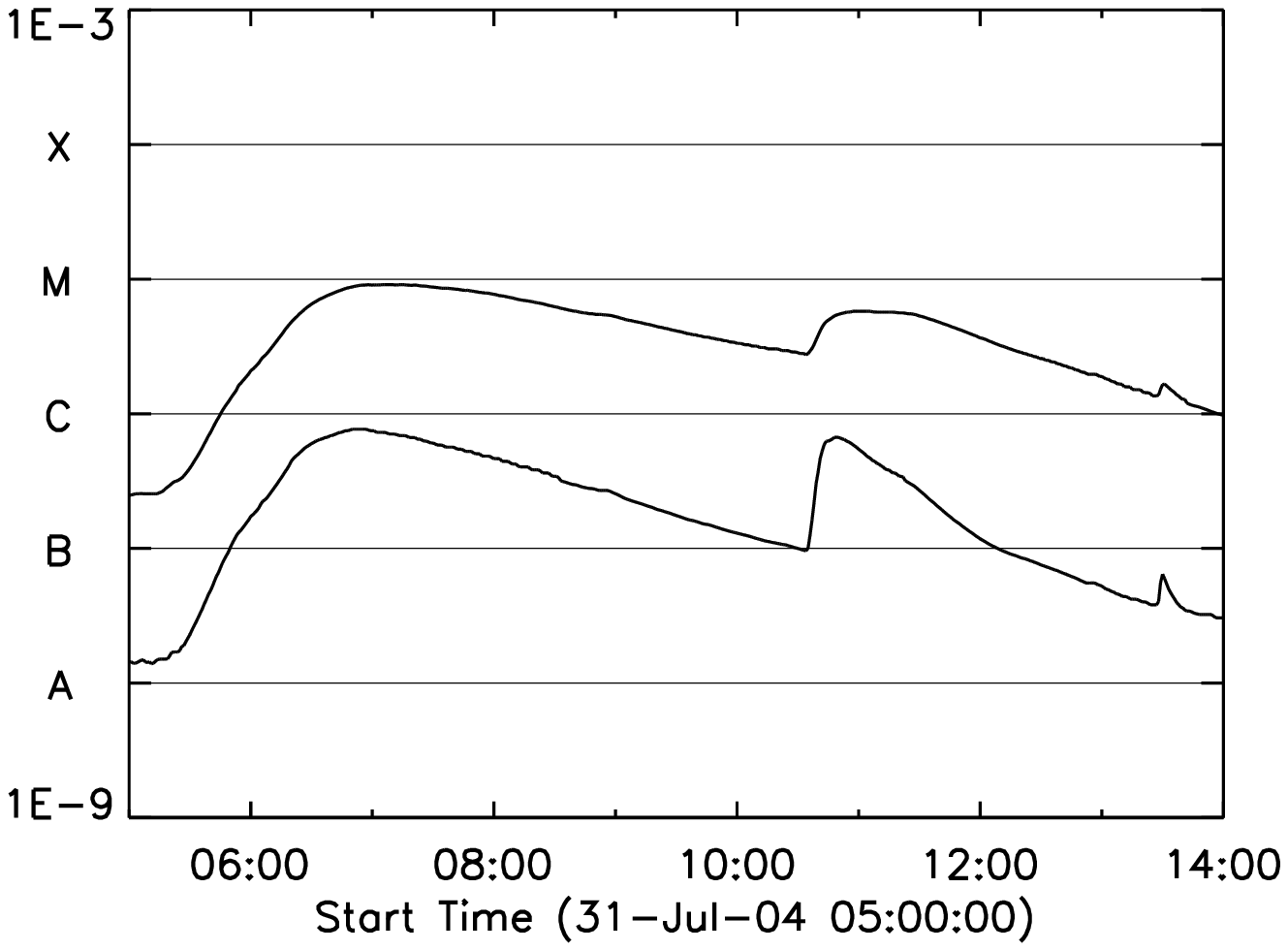}
\includegraphics[width=5.3cm]{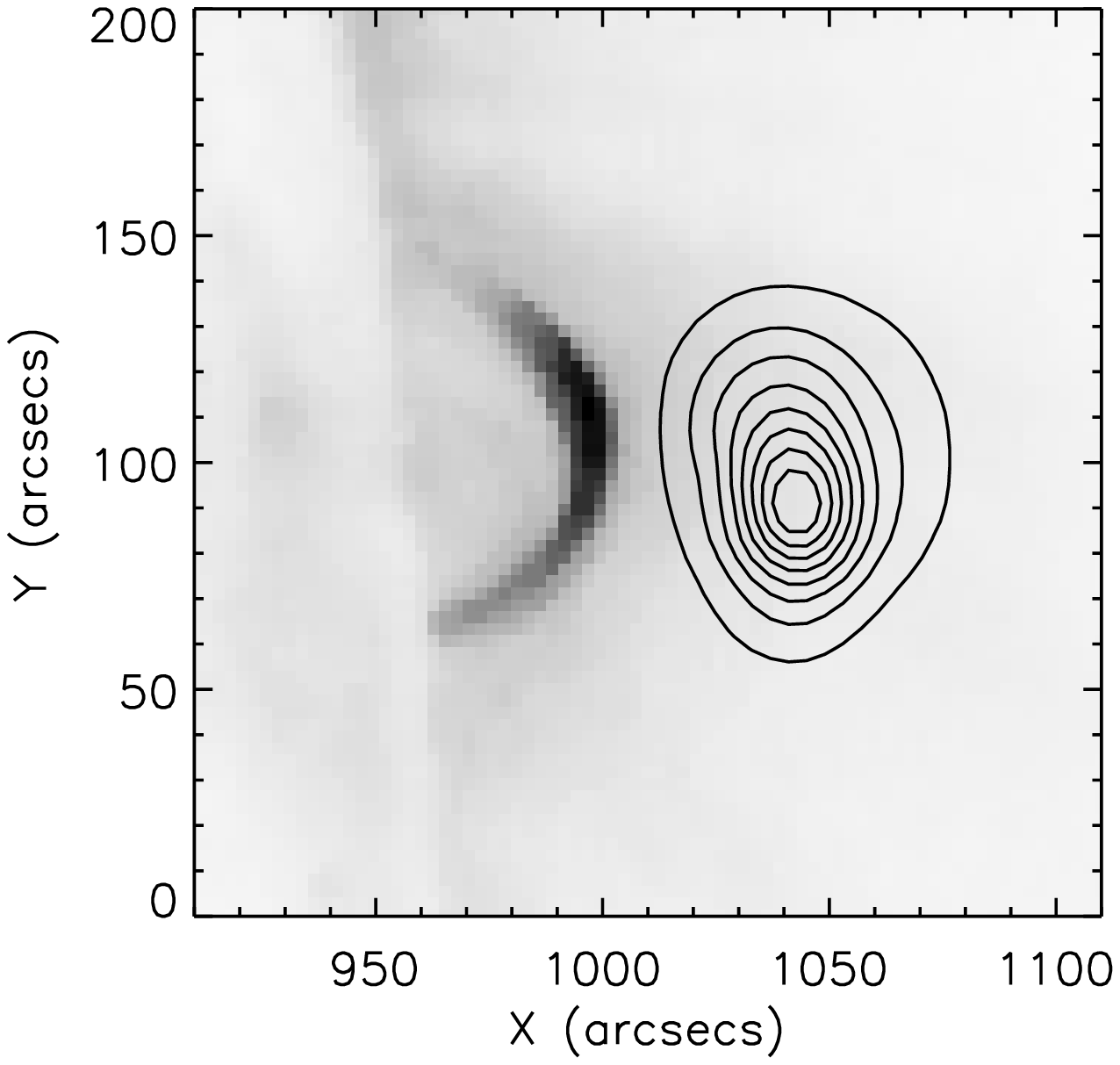}
\caption{{\em Left:}{\em GOES} X-ray fluxes (upper curve: $1-8$ \AA, lower curve: $0.5-4$ \AA). {\em Right:} {\em SoHO}/EIT 171 \AA $\ $ image showing the flare of 2004 July 31 during the rise phase. Contours show the emission in the $8-9$~keV range observed with {\it RHESSI}. The contours are for 10\% to 90\% (with increment 10 \%) of maximum emission.}
\label{goes31jul04}
\end{center}
\end{figure}

\begin{figure}[!h]
\begin{center}
\includegraphics[width=12.6cm]{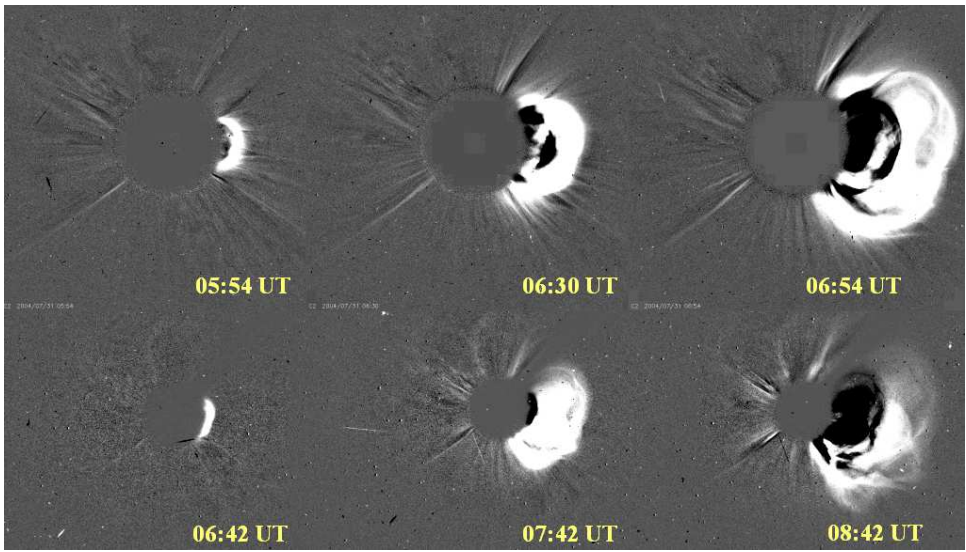}
\caption{Running-difference images of LASCO/C2 ({\em upper panels}) and LASCO/C3 ({\em lower panels}) illustrating the evolution of the 2004 July 31 CME.}
\label{lasco31jul04}
\end{center}
\end{figure}
\subsection{2004 July 31 event}
The CME was associated with the slow LDE, which occurred in the Active Region NOAA 10652, $\sim15^\circ$ behind the western solar limb. Soft X-ray (SXR) flux started to increase at 05:16 UT and reached its maximum ({\em GOES} class C8.4) at 06:57 UT (Fig. \ref{goes31jul04}, left panel). The flare was observed by many instruments: {\em SoHO}/EIT (after 05:24 UT), {\em GOES}/SXI and {\em RHESSI} (between 06:30--07:15 UT).

The Hard X-ray (HXR) emission was observed above a loop seen in the EIT images. HXR source was observed at high altitude, $h>80$ Mm, (Fig. \ref{goes31jul04}, right panel).

The CME was observed by LASCO/C2 and C3 coronographs (Fig. \ref{lasco31jul04}). Unfortunately, the EIT telescope observed the CME after 05:24 UT, 8 minutes after the flare start, so it was not possible to analyse CME at the very beginning. A leading edge of the CME was visible in the C2 images during at least 1 hour. Over the next few hours CME was in the FOV of the C3 coronograph.
 
\begin{figure}
\begin{center}
\includegraphics[width=6.0cm]{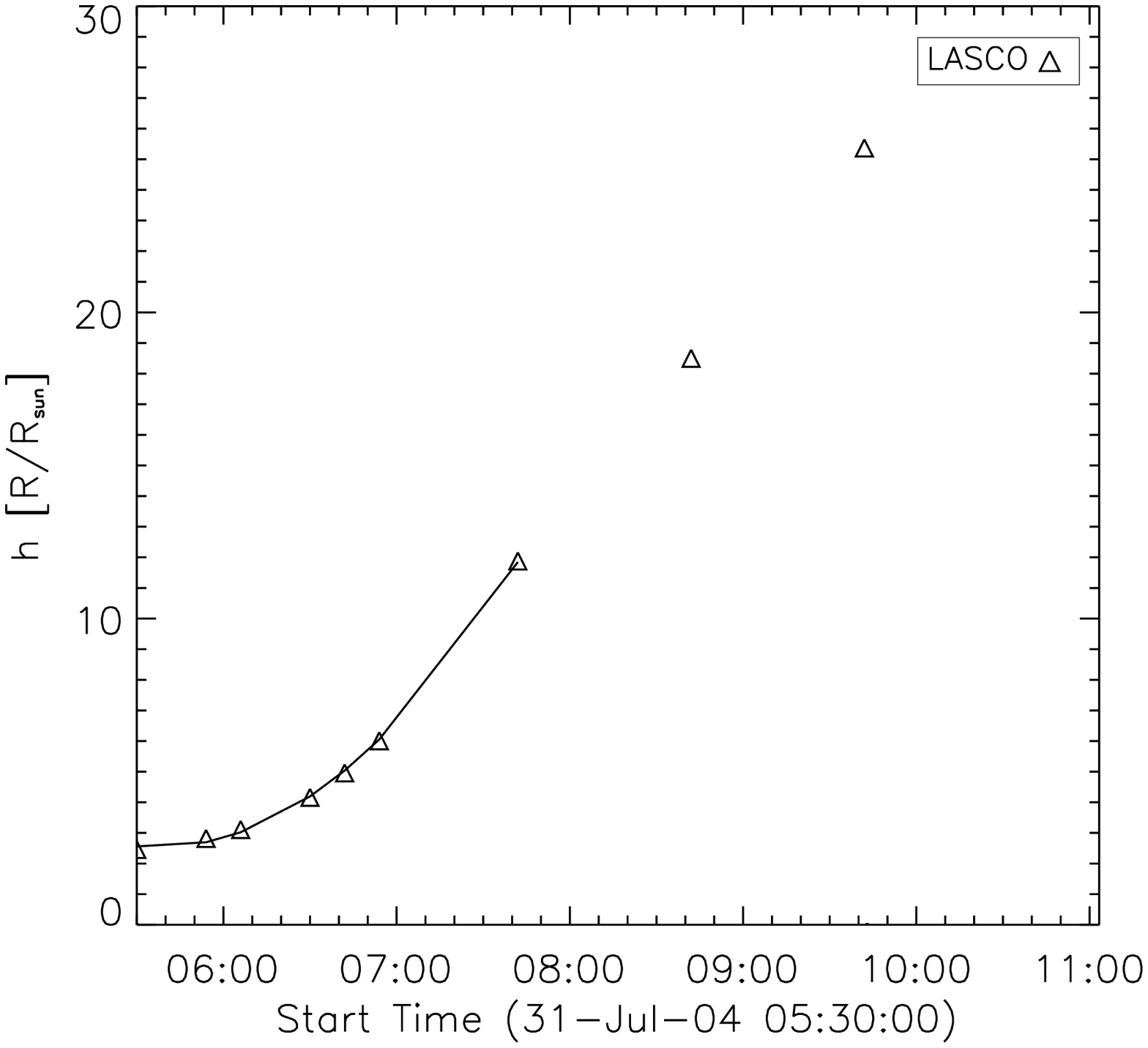}
\includegraphics[width=6.2cm]{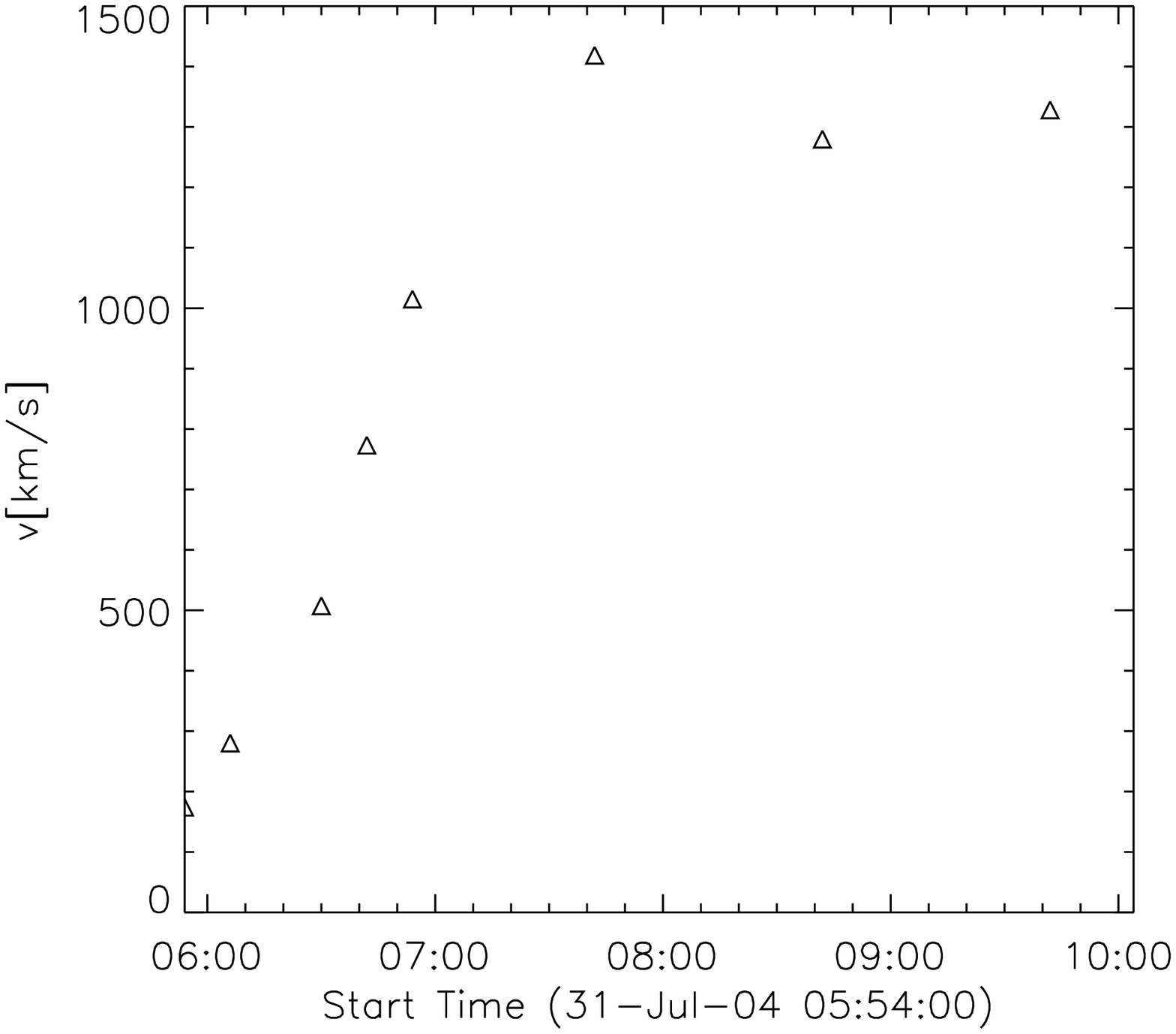}
\caption{{\em Left:} Height-time profile of the 2004 July 31 CME leading edge. Solid line shows second-degree polynomial fitting to the points, which was used to calculate average acceleration during the main acceleration phase. {\em Right}: Velocity-time profile of the CME leading edge.}
\label{ht_v31jul04}
\end{center}
\end{figure}

We measured the height of the CME (Fig. \ref{ht_v31jul04}, left panel) and used it to calculate an average acceleration during the main acceleration phase and a velocity profile (Fig. \ref{ht_v31jul04}, right panel). Velocity increased until $\sim$07:40 UT and reached $v_{max}$~$\sim$~$1400$ km/s (the average velocity during the propagation phase was about $1300$ km/s). The velocity profile is similar to the {\em GOES} light curve, which suggests a connection between the CME and the flare. The main acceleration phase was characterized by the average acceleration of about $230$~m/s$^2$. If we take into account the poor temporal resolution of the observations, we can assume that the CME is accelerated to the height  $\sim$~$6-12$~$R_\odot$. This value is greater than a value during a typical impulsive CME. The high velocity of the CME is caused by low, but prolongated acceleration, which lasted more than 2 hours.

\subsection{2005 September 6 event}
\begin{figure}
\begin{center}
\includegraphics[width=6.7cm]{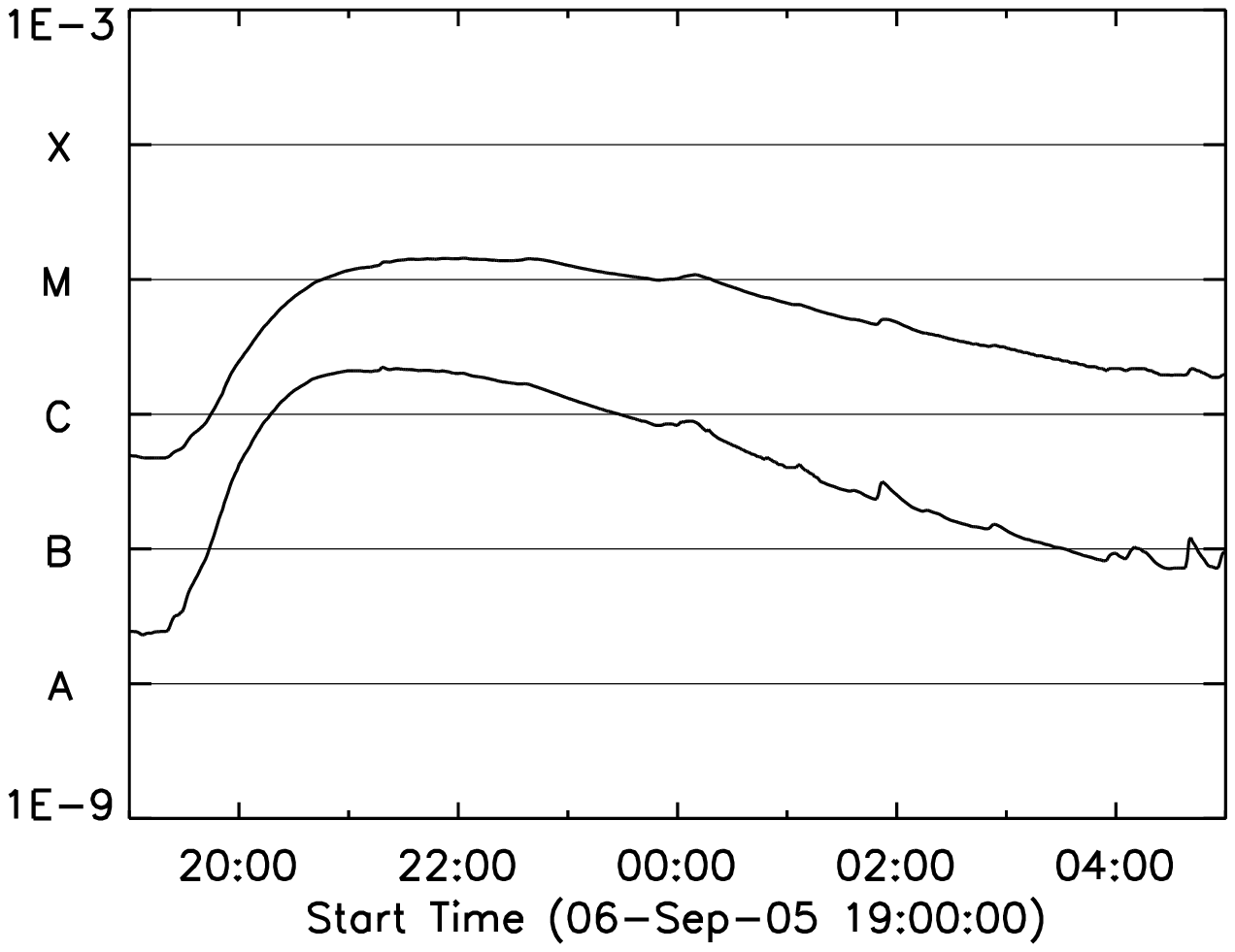}
\includegraphics[width=5.5cm]{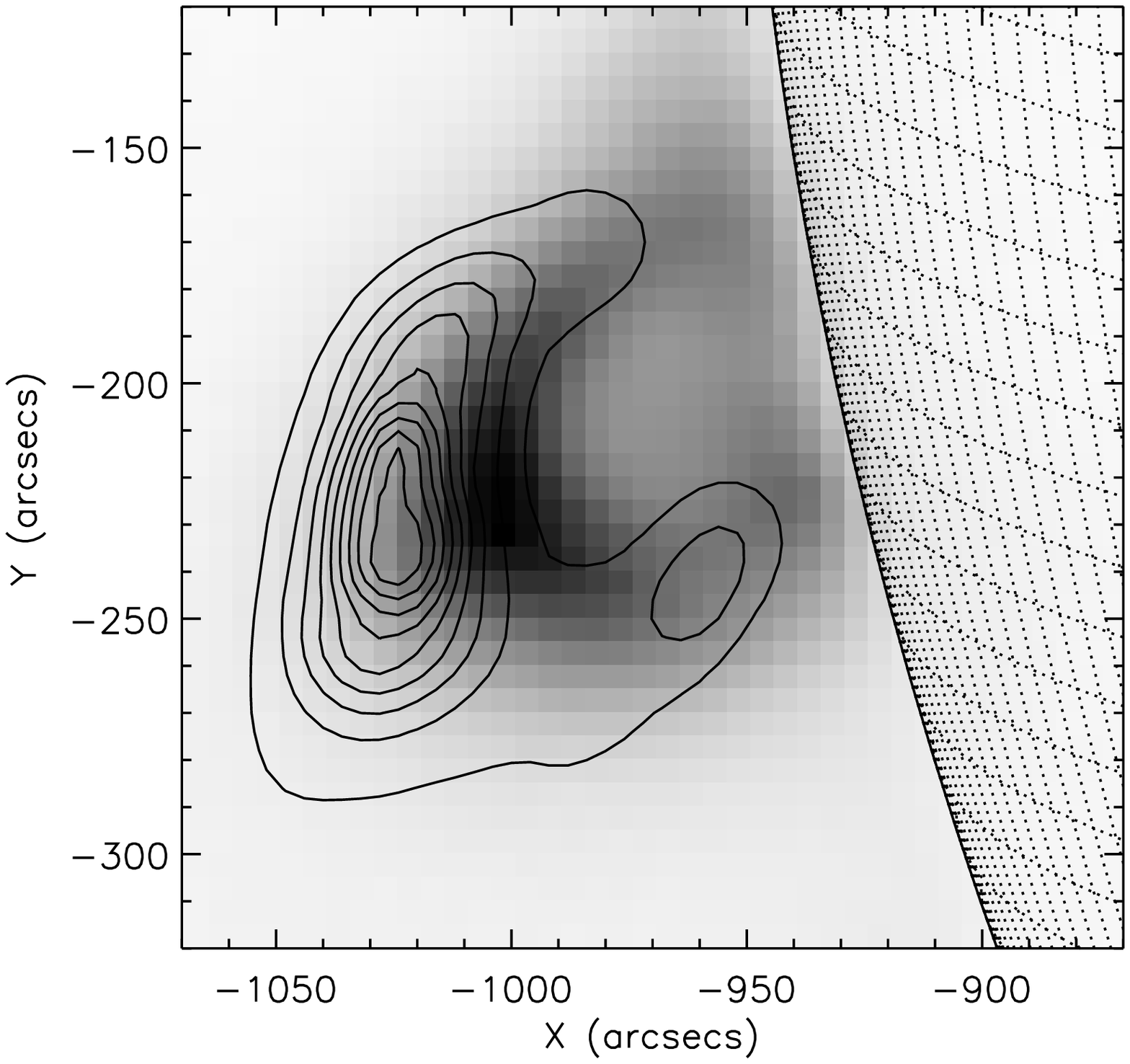}
\caption{{\em Left:}{\em GOES} X-ray fluxes (upper curve: $1-8$ \AA, lower curve: $0.5-4$ \AA). {\em Right:} {\em GOES}/SXI image showing flare of the 2005 September 6 during the rise phase. Contours show the emission in the $8-9$~keV range observed with {\it RHESSI}. The contours are for 10\% to 90\% (with increment 10 \%) of maximum emission.}
\label{goes6sep05}
\end{center}
\end{figure}

The 2005 September 6 CME was associated with a flare that occurred in the Active Region NOAA 10808, $\sim 6^\circ$ behind the eastern limb. The flare began at 19:32 UT and reached its maximum ({\em GOES} class M1.4) at 22.02 UT (Fig. \ref{goes6sep05}, left panel). The rise phase of the flare lasted $150$ minutes. The flare was well observed by {\em RHESSI} and {\em GOES}/SXI instruments. The HXR emission source was observed above the loop seen in the SXI images (Fig. \ref{goes6sep05}, right panel). A loop-top source seen in {\em RHESSI} images was observed at the height $h>70$~Mm and had size $r\sim16-20$ Mm. Only thermal emission was observed, however footpoints were occulted. The detailed analysis of this flare \citep{bak2010} revealed that thermal energy was released very slowly and decreased extremely slow after reaching the maximum value (characteristic time $\tau$ of heating rate decrease was $4430$ s) which can explain long rise phase. Did the energy release slowly also in the associated CME? 

\begin{figure}[!t]
\begin{center}
\includegraphics[width=12.6cm]{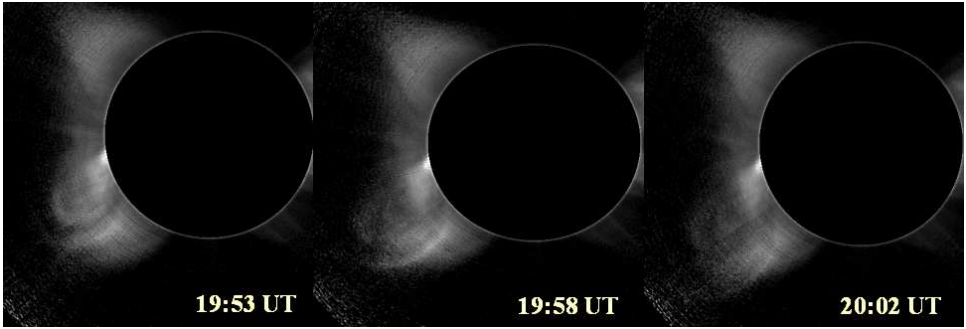}
\caption{MLSO/MK4 white light images illustrating the early stage of the 2005 September 6 CME evolution.}
\label{mk4}
\end{center}
\end{figure}

\begin{figure}[!t]
\begin{center}
\includegraphics[width=12.6cm]{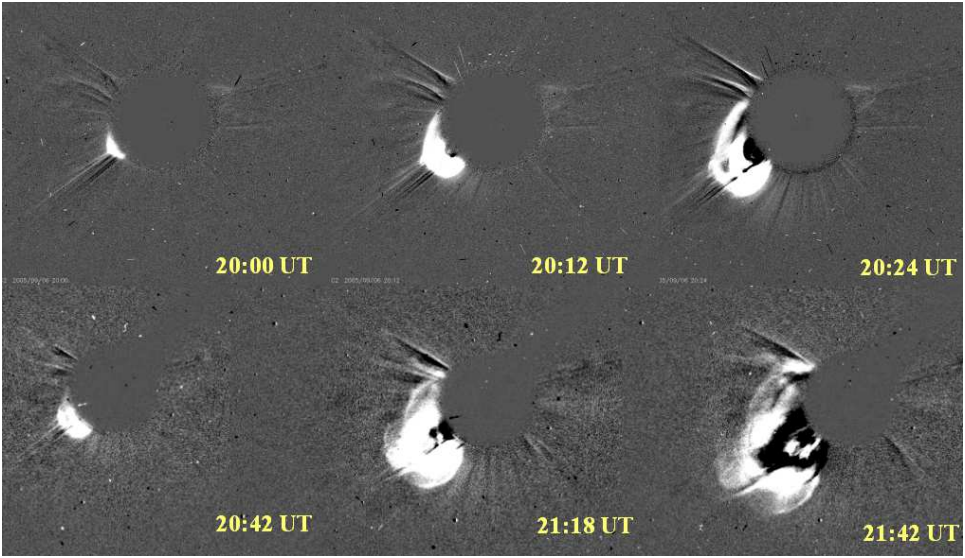}
\caption{Running-difference images of LASCO/C2 ({\em upper panels}) and LASCO/C3 ({\em lower panels}) illustrating the evolution of the 2005 September 6 CME.}
\label{lasco6sep05}
\end{center}
\end{figure}

\begin{figure}[!h]
\begin{center}
\includegraphics[width=6cm]{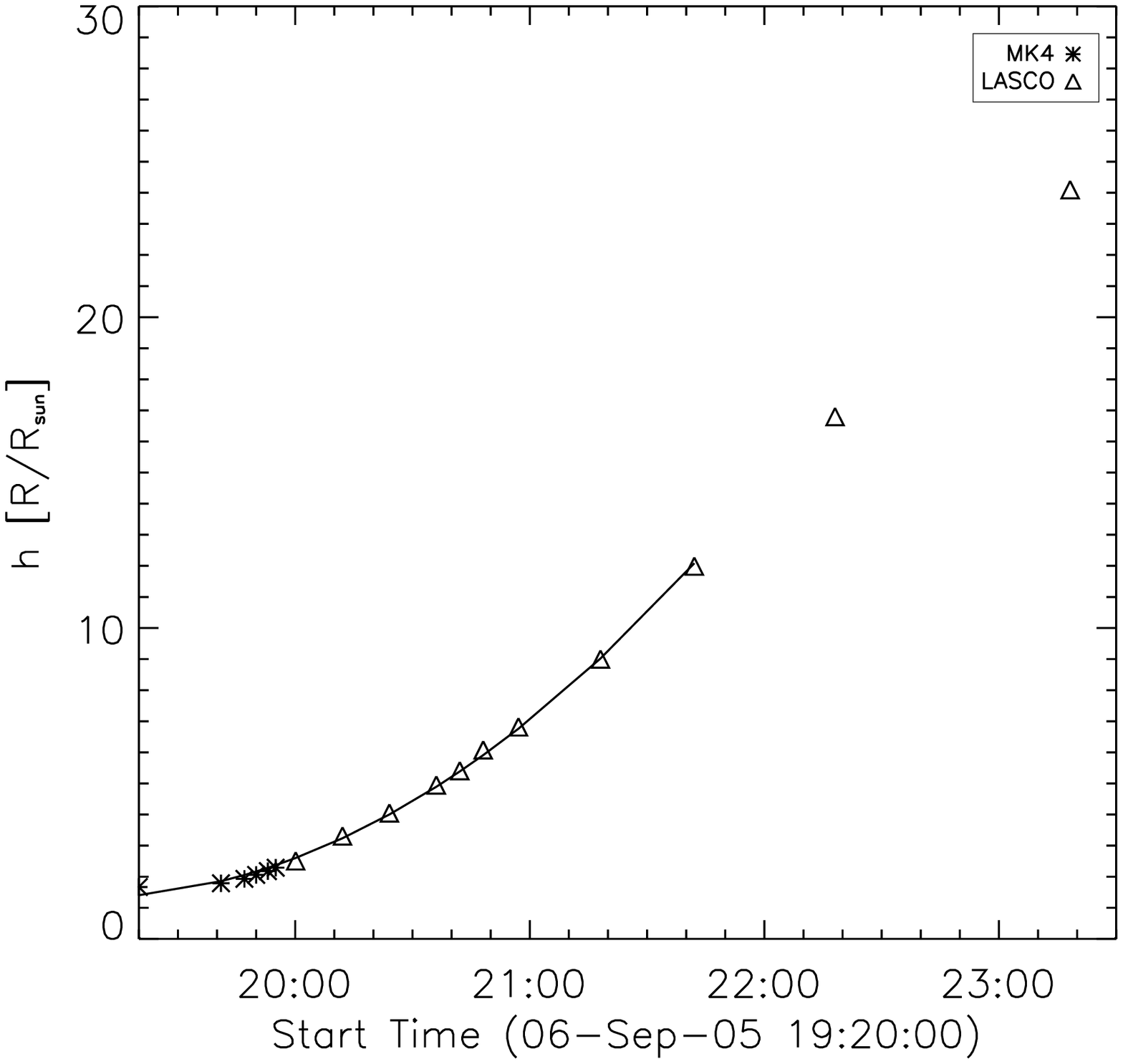}
\includegraphics[width=6.2cm]{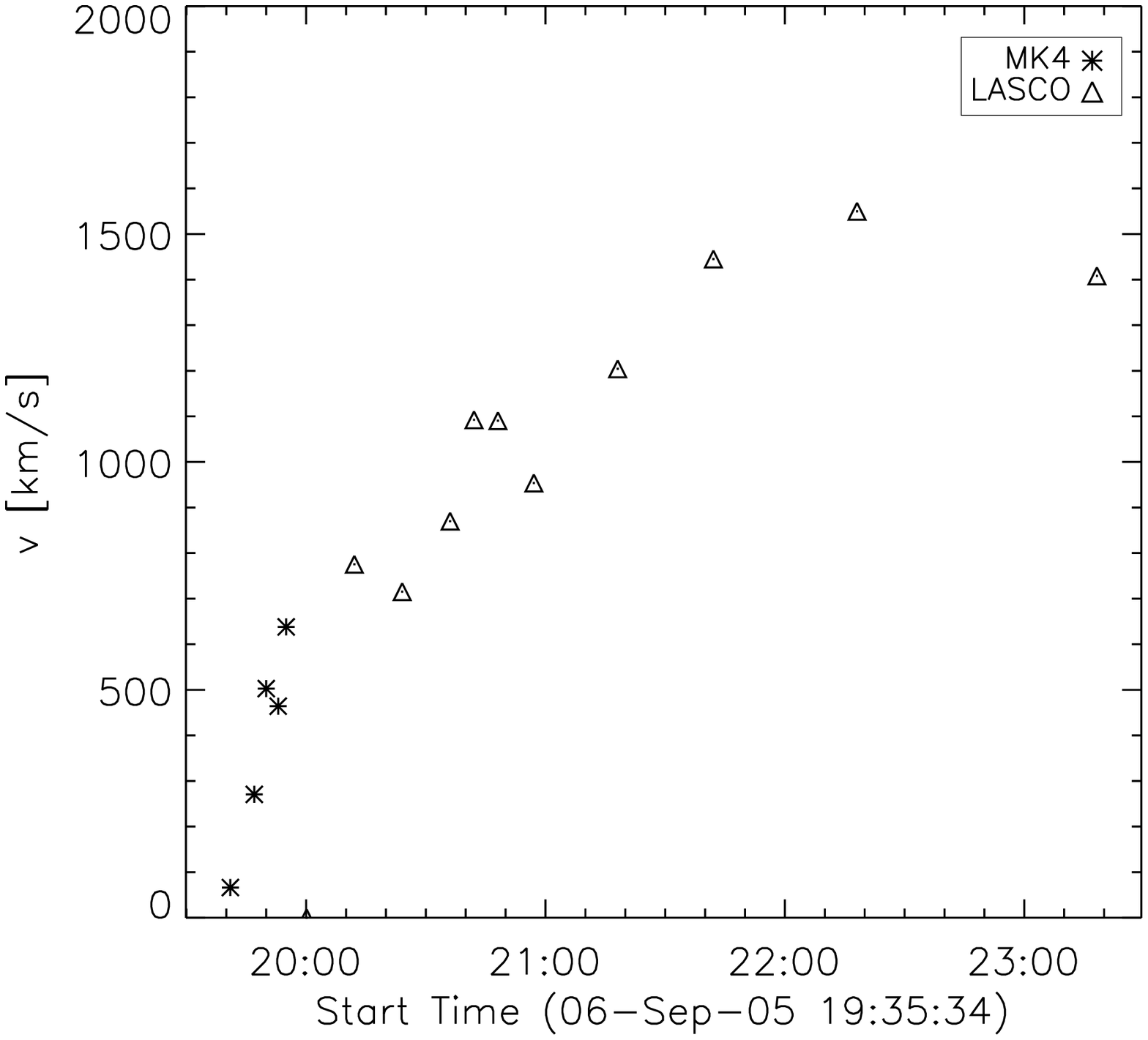}
\caption{{\em Left:} Height-time profile of the 2005 September 6 CME leading edge. Solid line shows second-degree polynomial fitting to the points, which was used to calculate average acceleration during the main acceleration phase.{\em Right}: Velocity-time profile of the CME leading edge.}
\label{ht_v6sep05}
\end{center}
\end{figure}

The CME was observed by MLSO MK4 (Fig. \ref{mk4}) and LASCO/C2 and C3 coronographs (Fig. \ref{lasco6sep05}). MK4 observations allowed us to investigate early stage of the evolution. Using all available observations, we measured height of the CME (h-t profile is shown in Fig. \ref{ht_v6sep05}, left panel). Velocity increased until $\sim$ 22:00 UT (Fig. \ref{ht_v6sep05}, right panel), to the flare maximum. To that time CME was accelerated with the average value of $170$~m/s$^2$. Average velocity during the propagation phase was equal to $1460$~km/s, maximum value was greater than $1500$~km/s. This high value is due to the very long acceleration phase, which lasted more than 2 hours. The CME was accelerated up to the height $\sim$~$12-17$~$R_\odot$, which is a much greater value than a value in a typical CME. Height-time profile is similar to the profile for a CME associated with an eruptive prominence, however the CME was associated with the flare, in which slow energy release was also observed.

\section{Results and summary}
We analysed five CMEs associated with slow LDE flares. We measured heights of the CMEs, estimated acceleration phase duration and calculated average velocity during the propagation phase and average acceleration during the main acceleration phase. The results for all the CMEs are given in Table \ref{CMEs}. 

\begin{table}[!t]
\begin{center}
	\caption{Characteristics of analysed CMEs}
	\vspace{3mm}
	\label{CMEs}
	\begin{footnotesize}
		\begin{tabular}{|c|c|c|c|c|c|}
\hline
&Flare rise&Acceleration&Average&Average&\\
Date&phase duration&phase duration$^{1}$&velocity$^{2}$&acceleration$^{3}$& Height$^{4}$\\
&[min]&[min]&[km/s]&[m/s$^2$]&[R$_\odot$]\\
\hline
21-May-02&76&78&1250&280&5.2--6.9\\
\hline
28-Jul-04&217&168&900&94&7.4--10.0\\
\hline
29-Jul-04&82&108&1400&140&4.8--9.3\\
\hline
31-Jul-04&101&132&1300&230&6.0--11.9\\
\hline
06-Sep-05&150&178&1460&170&12.0--16.80\\
\hline
\multicolumn{6}{@{} l @{}}{$^{1}$ time range used for quadratic fitting} \\
\multicolumn{6}{@{} l @{}}{$^{2}$ during the propagation phase} \\	
\multicolumn{6}{@{} l @{}}{$^{3}$ during the acceleration phase} \\
\multicolumn{6}{@{} l @{}}{$^{4}$ to which the CME was accelerated} \\
	\end{tabular}
	\end{footnotesize}
\end{center}
\end{table}
We can summarize our analysis as follows:
\begin{itemize}
\item CMEs associated with slow LDE flares are characterized by high velocity (in most cases $v>1000$ km/s). This high velocity was caused by low, but prolongated acceleration. 
\item Duration of the acceleration phase was longer than during typical CMEs. Duration of this phase is related to duration of the rising phase of associated flares. 
\item Average value of acceleration during the main acceleration phase was low ($a<300$ m/s$^2$ for analysed CMEs). We obtained the lowest value of acceleration for the CME connected with the flare with the slowest rise phase. 
\item CMEs were accelerated to the heights $h\sim 5$--$17R_\odot$ (Table \ref{CMEs}), which is greater value than value in typical impulsive CMEs.
\item All analysed CMEs were accelerated for at least 1 hour. This prolongated acceleration caused that the h-t profiles are more similar to CMEs associated with eruptive prominences, although they are connected with the slow LDE flares and their evolution can be described in three-phase scenario. Our results regarding this type of events indicate that there is no clear distinction flare-associated and prominences-associated CMEs. 

\end{itemize}

\section*{Acknowledgements} 
We thank {\em RHESSI}, {\em SoHO}, {\em GOES} and MLSO teams for their open data policy. We thank the referee for useful comments and suggestions. We also thank Barbara Cader-Sroka for editorial remarks. This investigation has been supported by a Polish Ministry of Science and High Education, grant No. N203 1937 33.

\end{document}